\documentclass[pop-mod,fleqn]{w-art-mod}
\usepackage{times}
\usepackage{w-thm}
\usepackage{amsmath}
\usepackage{amsfonts}
\usepackage{amssymb}

\usepackage[american]{babel}
\usepackage{graphicx}
\begin{document}
\pagespan{1}{}




\title[Fixing Moduli in Exact Type IIA Flux Vacua]{Fixing Moduli in Exact Type IIA Flux Vacua}


\author[F. Benini]{Francesco Benini\inst{1,}%
  \footnote{\quad E-mail:~\textsf{benini@sissa.it}}}
\address[\inst{1}]{International School for Advanced Studies (SISSA/ISAS), Via Beirut 2-4, 34014 Trieste, ITALIA \\
and INFN - Sezione di Trieste.}
\begin{abstract}
Type IIA flux compactifications with O6-planes have been argued from a four dimensional
effective theory point of view to admit stable, moduli free solutions. We discuss in detail the
ten dimensional description of such vacua and present exact supersymmetric solutions in the case when the
O6-charge is smoothly distributed: the geometry is Calabi-Yau and the dilaton is constant. In the localized case, the solution is a half-flat, non-Calabi-Yau
metric but still with constant dilaton. Finally, using the ten dimensional description we show how all moduli are stabilized 
and reproduce precisely the results of de Wolfe et {\it al.} \cite{DeWolfe:2005uu}. Based on \cite{Acharya:2006ne}.

\bigskip\bigskip
\noindent\emph{This article is based on a talk given by the author at the RTN Workshop ''Constituents, Fundamental Forces and Symmetries of the Universe'', Naples, Italy, 9-13 Oct 2006.}
\end{abstract}
\maketitle                   






\newcommand{\hyph}[1]{$#1$\nobreakdash-\hspace{0pt}}
\newcommand{\Nugual}[1]{$\mathcal{N}= #1 $}
\newcommand{\calO}{\mathcal{O}}
\newcommand{\calW}{\mathcal{W}}
\newcommand{\re}{\,\mathbb{R}\mbox{e}\,}
\newcommand{\im}{\,\mathbb{I}\mbox{m}\,}
\newcommand{\vol}{\mathrm{Vol}}
\newcommand{\eg}{{\it e.g.} }

\section{Introduction}

String vacua with magnetic fields turned on in the compact extra dimensions (``flux compactifications'') have attracted much attention since many years. The crucial feature is that their contribution to the energy of the system depends on the moduli of the geometry which supports them, so to provide an effective four dimensional potential for the moduli fields, possibly lifting some or all of them. Moreover the fluxes are quantized, so their strength is not another continuously varying free parameter. Together these give a mechanism for moduli stabilization.

From this point of view, particularly interesting are Type IIA vacua, since in many models a complete stabilization of the moduli can be accomplished at tree level in supergravity, that is without appealing to $\alpha'$ corrections or actually to string loops or non-perturbative effects. This is because the classical effective moduli potential generated by the tree level IIA supergravity action (supplemented with O6-plane sources) has stable \emph{isolated} minima \cite{DeWolfe:2005uu,Grimm:2004ua,Villadoro:2005cu}. 

Being concrete \cite{DeWolfe:2005uu,Villadoro:2005cu}, if we consider Type IIA theory on a Calabi-Yau three-fold, switching on R-R fluxes $F_0$, $F_2$, $G_4$ stabilizes all the K\"ahler moduli of the internal manifold. To stabilize also the complex structure moduli, one has to introduce a NS-NS 3-form flux $H_3$. However this gives a tadpole for the D6-charge, that can be cancelled by the introduction of orientifold six-planes. They also help in halving the complex structure moduli. The full system then stabilizes all the moduli, essentially at leading order in $\alpha'$ and $g_s$.

An approximation widely used in the literature to study string vacua in the supergravity regime, in which R-R and NS-NS field-strengths are switched on, is the so called ``Calabi-Yau with fluxes'' approximation. It resides in taking the flux contribution to the supergravity action and the equations of motion small compared to the curvature contribution, and it works in the large volume limit. Being the fluxes quantized by Dirac quantization condition, the minimum amount of a \hyph{p}form field-strength is $F_p \sim (\alpha')^{(p-1)/2}$. Then one requires the contribution of fluxes to the action to be small compared to the Einstein term $R$, which is of order $L^{-2}$ (with $L$ the characteristic length of the manifold). This gives $\alpha'/L^2 \ll 1$. In other words, we must be in the limit of large compactification manifold with respect to the string length, which anyway is the regime of applicability of supergravity. Under these conditions, one can neglect the backreaction of fluxes on the geometry, and work with Calabi-Yau metric solutions. Of course one has to be careful to remember that in the action the various contributions are weighted by the string coupling, and both the dilaton and the volume are (possibly) determined by the fluxes themselves, so it is not always possible to keep the fluxes to their minimal amount while increasing the volume.

Such an approximation can also hide subtle phenomena. For example, even the minimum amount of flux allowed by Dirac quantization condition can, if correctly taken into account, change the topology of the manifold as well as cohomology classes and so on. It is not clear how the Calabi-Yau geometry can approximate disconnected topologies.

Another powerfull tool in the discussion of moduli stabilization is the reduction to a four dimensional effective theory, with the computation of the 4d effective potential for light fields. Then vacua are found by looking at its stationary points. Many issues are studied more easily than with a ten dimensional approach. For instance, stability is checked just by looking at the second derivatives of the potential. Moreover non-supersymmetric vacua can be analyzed as well, and there is some control on exotic vacua, such as non-geometric vacua \cite{NonGeometric}.

The approach has anyway also bad facets. First of all, one has to work in a Kaluza-Klein approximation, expanding the ten dimensional fields on a Kaluza-Klein basis (suitable for a Calabi-Yau geometry without fluxes) and then keeping only light fields. In general there is no guaranty that the same basis furnishes the light fields in presence of background fluxes.%
\footnote{For some attempts to do the Kaluza-Klein reduction with fluxes see \cite{KKReduction,House:2005yc}.}
Moreover, even finding stable minima of the 4d effective potential, one cannot be in general sure about the existence of a 10d lift. All these reasons brought us to undertake the ten dimensional approach, and the following discussion is based on \cite{Acharya:2006ne}.

In \cite{DeWolfe:2005uu} Type IIA string vacua in the supergravity limit were constructed, by compactifying on the orbifold $T^6/(\mathbb{Z}_3)^2$ and then adding O6-planes. Such vacua were analyzed classically and in the large volume regime, by computing the 4d effective potential and deriving it from a superpotential. Our main result will be to provide a full 10d description of their vacua with backreacting fluxes \cite{Acharya:2006ne}, reproducing their results.

A key ingredient is a smearing procedure. Indeed we show that in no case with non-trivial fluxes and localized sources (O6-planes) the geometry is Calabi-Yau, and this makes it hard to exhibit explicit examples. We overcome the problem by homogeneously distributing the O6-charge on the internal manifold, which makes it possible to find Calabi-Yau solutions with backreacting non-trivial fluxes. We stress that all these solutions have \emph{constant} dilaton, and the string coupling can be kept arbitrarily small everywhere.

Another approach attempted to give a full description of the vacua in 
\cite{DeWolfe:2005uu} appeared in \cite{Banks:2006hg}. The authors performed a double T-duality of the system to get rid of the $F_0$ and $H_3$ fluxes, and then uplifted to M-theory.

\section{Type IIA Supergravity Solutions with O6-planes}

The starting point to give a full ten dimensional description in the supergravity regime of the vacua constructed in \cite{DeWolfe:2005uu} are the Lust-Tsimpis solutions \cite{Lust:2004ig} (see also \cite{Behrndt:2004km}). They are the most general solutions of Type IIA supergravity compactified on a warped product of $AdS_4$ and an internal \hyph{SU(3)}structure manifold, which preserve \Nugual{1} supersymmetry in 4d. All the fluxes compatible with the Poincar\'e symmetry of $AdS_4$ are switched on. These include also a ``mass parameter'' $F_0 \equiv m$, which is one of the R-R fluxes provided by IIA string theory and at low energy is described within the so called massive Type IIA supergravity \cite{Romans:1985tz}. Then we will add O6-planes to them in a way to preserve \Nugual{1} supersymmetry, in order to be able to realize, among the others, the $T^6/(\mathbb{Z}_3)^2$ orientifold model of \cite{DeWolfe:2005uu}.

The internal manifold is taken to be an \hyph{SU(3)}structure manifold, so to have one globally defined spinor that, combined with the four dimensional ones, lets us construct four 10d Killing spinors. An \hyph{SU(3)}structure manifold is described by a K\"ahler 2-form $J$ and an holomorphic 3-form $\Omega$, like on a Calabi-Yau 3-fold. But, unlike that case, they are not closed. Rather their differentials, conveniently split according to $SU(3)$ representations, define the torsion classes and measure the deviation from the Calabi-Yau geometry \cite{Chiossi:2002tw}:
\begin{align}
dJ &= -\frac{3}{2} \im \bigl( \calW_1^{1\oplus 1} \Omega^* \bigr) + \calW_4^{3\oplus \bar 3} \wedge J + \calW_3^{6\oplus\bar 6} \\
d\Omega &= \calW_1^{1\oplus 1} J\wedge J + \calW_2^{8\oplus 8} \wedge J + \calW_5^{3\oplus\bar 3}\wedge \Omega \;.
\end{align}
Here $\calW_i$ are the five torsion classes, and the specific $SU(3)$ representation is indicated.

In the present setup, supersymmetry only allows the non-vanishing torsion classes $\calW_1^-$ and $\calW_2^-$ \cite{Lust:2004ig}, where minus stands for the imaginary part. This restricts the possible geometries to a special subclass of \emph{half-flat} manifolds.

We can introduce orientifold six-planes in the picture. These are not genuine supergravity objects, rather they are defined by projecting the Type IIA string theory. However the supergravity action can be enriched with terms that describe the interaction of O-planes with the low energy fields.
In Type IIA string theory, an O6-plane is obtained by modding out the theory by the discrete symmetry operator $\calO = \Omega_p (-1)^{F_L} \sigma^*$, where $\Omega_p$ is the world-sheet parity, $F_L$ is the left-moving spacetime fermion number and $\sigma$ is an isometrical involution of the background. The fixed point locus of $\sigma$ is the O6-plane. In Type IIA string theory an O6-plane is a BPS object that preserves half of the supersymmetries, by relating the two Majorana-Weyl supersymmetry parameters according to $\epsilon_\pm = \calO \epsilon_\mp$. In the vacua we are going to construct we want to preserve 4d Poincar\'e symmetry, so the O6-planes will span the $AdS_4$ factor and will wrap a three dimensional cycle in the internal manifold. In order to obtain again vacua with four supercharges, we must place the O6-planes in such a way to preserve the four supercharges of the background. This is achieved by wrapping the planes on supersymmetric 3-cycles $\Sigma$, which in the case of \hyph{SU(3)}structure manifolds are generalized special Lagrangian 3-cycles%
\footnote{Supersymmetry forces $\sigma$ to be antiholomorphic with respect to the almost complex structure J, and the pull-back of the NS-NS 2-form potential $B_2$ on $\Sigma$ to vanish.}
\cite{Cascales:2004qp}:
\begin{equation}
J|_\Sigma =0 \qquad \qquad
\re\Omega|_\Sigma =0 \qquad \qquad
\im\Omega|_\Sigma = -\vol_\Sigma \;.
\end{equation}

The description of the low energy dynamics is achieved, at leading order in $\alpha'$, by adding to the Type IIA supergravity action a Dirac-Born-Infeld and a Wess-Zumino term for the O6-planes. The main modification is a violation of the Bianchi identity for the R-R 2-form field-strength $F_2$, since O6-planes couple to the R-R 7-form potential $C_7$ and are thus magnetic sources for $F_2$:
\begin{equation}
dF_2 = 2mH_3 - 2\mu_6 \, \delta_3 \;,
\end{equation}
where $\delta_3$ is the O6-charge distribution, a 3-form localized on $\Sigma$. $H_3$ is the NS-NS 3-form field-strength and $\mu_6$ is the Einstein frame D6-charge. By imposing supersymmetry we can find the following solutions, where everything is determined by the geometry:
\begin{align} \label{IIA solution}
F_2 &= \frac{f}{9} \, e^{-\phi/2} J + \tilde{F_2} &
H_3 &= \frac{4m}{5} \, e^{7\phi/4} \, \re \Omega &
-\frac{4i}{9} f \, e^{\phi/4} &= \calW_1^- \nonumber \\
G_4 &= f \, dVol_4 + \frac{3m}{5} e^{\phi} J\wedge J  &
\phi \,,\, & f \,,\, \Delta = \mbox{constant} &
-i \, e^{3\phi/4} \tilde{F_2} &=\calW_2^- \;.
\end{align}
In particular, the geometry is again a special subclass of half-flat.
Some explanations are in order. $G_4$ is the R-R 4-form field-strength, $f$ is a Freund-Rubin parameter, $\phi$ is the IIA dilaton and $\Delta$ is the warp factor which is found to be constant. 
$\tilde{F_2}$ is the irreducible component of $F_2$ transforming in the $\mathbf{8}$ of $SU(3)$, and is constrained by the O6-charge distribution:
\begin{equation} \label{tilde F2}
d\tilde{F_2} = - \frac{2}{27} e^{-\phi/4} \Bigl( f^2 - \frac{108}{5} m^2 e^{2\phi} \Bigr)
\re\Omega - 2 \mu_6 \, \delta_3 \:.
\end{equation}
Generalizing the argument in \cite{Lust:2004ig} one can show that these are actually solutions of the equations of motion. 

\

From \eqref{IIA solution} and \eqref{tilde F2}, keeping $\delta_3$ a localized form, we can derive the inequality $f^2 \geq \frac{108}{5} m^2 e^{2\phi}$, from which an important observation follows. As long as the mass parameter $m$ is different from zero, $\calW_1^-$ is non-trivial and there are \emph{no Calabi-Yau} solutions (by careful analysis, $m=0$ leads to trivial Calabi-Yau solutions without fluxes and O6-planes). Constructing explicit examples of half flat manifolds which solve the equations, even without O6-planes, is difficult (for some attempts see \cite{House:2005yc,Lust:2004ig}). One possible way out is to apply a smearing procedure, much like the smearing of D-branes: the O6-charge is homogeneously distributed over the internal manifold, while the orientifold projection is kept untouched.%
\footnote{We must note that while the smearing of a large number of branes is perfectly justified, the smearing of a small number is less, as well as the fact that the orientifold projection cannot be smeared.}
By substituting the localized charge distribution $\delta_3$ with the smeared one:
\begin{equation} \label{smeared distrib}
\mu_6 \, \delta_3^{\text{smeared}} = \frac{4m^2}{5} \, e^{7\phi/4} \, \re\Omega
\end{equation}
it is possible to find solutions with $m^2>0$ and $f=0$, $F_2=\tilde{F_2}=0$. These are Calabi-Yau geometries with non-trivial backreacting fluxes.

More generally, every Calabi-Yau 3-fold that admits an anti-holomorphic involution provides a solution of our equations, with O6-planes along the fixed point locus of the involution. Moreover, as we will see in the next section, in such geometries all moduli can be stabilized.

By integrating \eqref{smeared distrib} on a cycle $\Gamma$ symplectic partner of $\Sigma$, it is possible to show that the dilaton is fixed to
\begin{equation}
e^{7\phi/4} = \frac{5\mu_6}{8m^2 \sqrt{\vol_6}} \;,
\end{equation}
and the 4d cosmological constant is fixed as well. Summarizing, the solution is completely described by an internal Calabi-Yau manifold defined by $J$ and $\Omega$, with an antiholomorphic isometrical involution $\sigma$. Further constraints come from the integral quantization of fluxes, and this mechanism provides the stabilization of \emph{all} moduli.

\section{Moduli Stabilization}

The next step is to use the class of solutions previously found to discuss the possible presence of free moduli. The question is -- is there a continuously connected family of vacua, or do they form only a discrete set? We will answer by exploiting the class of smeared Calabi-Yau solutions, but the same discussion could be carried over in the localized case.

First of all we split the VEV of a field strength in two pieces, \eg $H_3 = H_3^\text{flux} + dB_2$. The first piece, closed and non-trivial in cohomology, represents the integral flux contribution, while the second one, exact, encodes any other oscillation or deformation. In such a separation there is of course some arbitrariness. From our smeared Calabi-Yau solution, we know $H_3$ to be harmonic, so the natural choice is to take $H_3^\text{flux}$ harmonic as a representative of the cohomology class, and then use the gauge freedom to put also $B_2$ harmonic. The fact that this is always possible shows that cannot be any physical non-harmonic component in $B_2$: these have vanishing VEV and are massive. Then we can expand $H_3^\text{flux}$ and $B_2$ in an harmonic basis and substitute in the solution \eqref{IIA solution}, to see if free parameters are left.

\

We can easily handle the $T^6/(\mathbb{Z}_3)^2$ orientifold model of \cite{DeWolfe:2005uu}. The geometry is the orbifold limit of a Calabi-Yau manifold with $h^{2,1}=0$, $h^{1,1}=12$, where 9 of the 12 K\"ahler moduli arise from the blow-up modes of 9 $\mathbb{Z}_3$ singularities.
We will not discuss the twisted moduli here, since they fit well in the general Calabi-Yau case \cite{Acharya:2006ne}.
It's useful to introduce an integral basis of harmonic forms: the 2-forms (odd under $\sigma$) $w_i$ such that $\int w_1\wedge w_2\wedge w_3=1$ and $w_i^2=0$; the 4-forms (even under $\sigma$) $\tilde w^i = w_j \wedge w_k$, where $j,k$ are the two values among $1,2,3$ besides $i$. The field-strengths are split according to:
\begin{align}
F_2 &= F_2^\text{flux} +  dA_1 + 2 m\,B_2 \nonumber \\
H_3 &= H_3^\text{flux} + dB_2 \nonumber \\
G_4 &= G_4^\text{flux} + f \, dVol_4 + dC_3 + B_2\wedge dA_1 + mB_2^2 \nonumber \\
e^{\phi/2}\ast G_4 &= F_6^\text{flux} + dC_5 - H_3 \wedge C_3 + B_2 \wedge G_4^\text{flux} + \tfrac{1}{3}m\, B_2^3 \:.
\end{align}
Notice that being $A_1$ harmonic, it is actually vanishing on a Calabi-Yau manifold. Then we expand on the harmonic basis both the fluxes, integrally quantized, and the forms possibly containing free moduli:
\begin{align}
F_2^\text{flux} &= f^i \, w_i &
B_2 &= b^i \, w_i \\
G_4^\text{flux} &= e_i \, \tilde w^i &
J &= e^{-\phi/2} \, v^i \, w_i \\
H_3^\text{flux} &= \frac{p}{4\sqrt{\vol_6}} \, \re\Omega &
C_3 &= \frac{\xi}{4\sqrt{\vol_6}} \, \im\Omega \\
F_6^\text{flux} &= e_0 \, w_1\wedge w_2\wedge w_3 \;,
\end{align}
where $f^i$, $e_i$, $p$ and $e_0$ (as well as $m$) are quantized in suitable units, and $\vol_6= e^{-3\phi/2} \, v^1 v^2 v^3$. The could-be moduli are $v^i$, $\phi$ and the axions $b^i$ and $\xi$, which pair into 4d chiral fields. The only non-trivial tadpole cancellation condition is $m \, p = \mu_6$. Substituting in the smeared solution we get $b^i = - f^i/2m$.
Taking for simplicity $f^i=0$ we also get:
\begin{equation}
v^i = \frac{1}{|e_i|} \sqrt{\frac{5}{6} \, \frac{e_1e_2e_3}{m}} \qquad \qquad
e^\phi = \frac{3}{4} \mu_6 \left( \frac{5}{6} \, \frac{1}{m^5\,e_1e_2e_3} \right)^{1/4} \qquad \qquad
p \, \xi = e_0 \;.
\end{equation}
We could also compute the 4d cosmological constant $\Lambda$ \cite{Acharya:2006ne}. Everything is in perfect agreement with the results of \cite{DeWolfe:2005uu}.

The same machinery applies to general Calabi-Yau solutions with general backreacting fluxes and smeared O6-planes, as discussed in \cite{Acharya:2006ne}. In general there are enough equations to fix almost all the would-be moduli: all the K\"ahler and complex structure moduli, the dilaton, all axions from $B_2$ and one axion from $C_3$. The remaining $C_3$ axions turn out not to be fixed yet, but this is not a phenomenological problem. Taking values on a compact $S^1$, there cannot be any dangerous runaway behavior and any other contribution to their effective potential lifts them.

\begin{acknowledgement}
FB warmly thanks the organizers of the RTN workshop ``Constituents, Fundamental Forces and Symmetries of the Universe'', 2006, for having given him the opportunity of present this talk.
\end{acknowledgement}


\begin{thebibliography}{[00]}

\bibitem{DeWolfe:2005uu}
  O.~DeWolfe, A.~Giryavets, S.~Kachru and W.~Taylor,
  JHEP {\bf 0507}, 066 (2005)
  [arXiv:hep-th/0505160].

\bibitem{Grimm:2004ua}
  T.~W.~Grimm and J.~Louis,
  Nucl.\ Phys.\  B {\bf 718}, 153 (2005)
  [arXiv:hep-th/0412277]. \\
  P.~G.~Camara, A.~Font and L.~E.~Ibanez,
  JHEP {\bf 0509}, 013 (2005)
  [arXiv:hep-th/0506066].

\bibitem{Villadoro:2005cu}
  G.~Villadoro and F.~Zwirner,
  JHEP {\bf 0506}, 047 (2005)
  [arXiv:hep-th/0503169].
  
\bibitem{NonGeometric}
  S.~Kachru, M.~B.~Schulz, P.~K.~Tripathy and S.~P.~Trivedi,
  JHEP {\bf 0303}, 061 (2003)
  [arXiv:hep-th/0211182]. \\
  S.~Hellerman, J.~McGreevy and B.~Williams,
  JHEP {\bf 0401}, 024 (2004)
  [arXiv:hep-th/0208174]. \\
  A.~Flournoy, B.~Wecht and B.~Williams,
  Nucl.\ Phys.\ B {\bf 706}, 127 (2005)
  [arXiv:hep-th/0404217]. \\
  J.~Shelton, W.~Taylor and B.~Wecht,
  JHEP {\bf 0510}, 085 (2005)
  [arXiv:hep-th/0508133]. \\
  A.~Lawrence, M.~B.~Schulz and B.~Wecht,
  JHEP {\bf 0607}, 038 (2006)
  [arXiv:hep-th/0602025].

\bibitem{KKReduction}
  S.~Gurrieri, J.~Louis, A.~Micu and D.~Waldram,
  Nucl.\ Phys.\ B {\bf 654}, 61 (2003)
  [arXiv:hep-th/0211102]. \\
  S.~Gurrieri and A.~Micu,
  Class.\ Quant.\ Grav.\  {\bf 20}, 2181 (2003)
  [arXiv:hep-th/0212278]. \\
  S.~Fidanza, R.~Minasian and A.~Tomasiello,
  Commun.\ Math.\ Phys.\  {\bf 254}, 401 (2005)
  [arXiv:hep-th/0311122]. \\
  A.~Tomasiello,
  JHEP {\bf 0506}, 067 (2005)
  [arXiv:hep-th/0502148]. \\
  M.~Grana, J.~Louis and D.~Waldram,
  JHEP {\bf 0601}, 008 (2006)
  [arXiv:hep-th/0505264]. \\
  A.~K.~Kashani-Poor and R.~Minasian,
  arXiv:hep-th/0611106.

\bibitem{House:2005yc}
  T.~House and E.~Palti,
  Phys.\ Rev.\ D {\bf 72}, 026004 (2005)
  [arXiv:hep-th/0505177].

\bibitem{Acharya:2006ne}
  B.~S.~Acharya, F.~Benini and R.~Valandro,
  JHEP {\bf 0702}, 018 (2007)
  [arXiv:hep-th/0607223].

\bibitem{Banks:2006hg}
  T.~Banks and K.~van den Broek,
  arXiv:hep-th/0611185.
  
\bibitem{Lust:2004ig}
  D.~Lust and D.~Tsimpis,
  JHEP {\bf 0502}, 027 (2005)
  [arXiv:hep-th/0412250].

\bibitem{Behrndt:2004km}
  K.~Behrndt and M.~Cvetic,
  Phys.\ Rev.\ Lett.\  {\bf 95}, 021601 (2005)
  [arXiv:hep-th/0403049]. \\
  K.~Behrndt and M.~Cvetic,
  Nucl.\ Phys.\ B {\bf 708}, 45 (2005)
  [arXiv:hep-th/0407263].

\bibitem{Romans:1985tz}
  L.~J.~Romans,
  Phys.\ Lett.\  B {\bf 169}, 374 (1986).

\bibitem{Chiossi:2002tw}
  S.~Chiossi and S.~Salamon,
  arXiv:math.dg/0202282.

\bibitem{Cascales:2004qp}
  J.~F.~G.~Cascales and A.~M.~Uranga,
  JHEP {\bf 0411}, 083 (2004)
  [arXiv:hep-th/0407132].





\end{thebibliography}
\end{document}